ORIGINAL PAPER

# When disclosure rules are not enough: authorial warrant and judgment in ai-supported scholarship

Marcin Mirończuk[1]

**Abstract**

Editorial policies increasingly require authors to disclose substantive uses of generative AI, but such requirements commonly focus on whether, where, and how a model was used. Even function-specific disclosure may locate a model in the research or writing process without showing how its outputs entered scholarly judgment, how they were evaluated, or whether contributors can reconstruct the relevant reasoning and defend resulting claims without treating those outputs as sufficient justification. Drawing on research ethics, AI disclosure policy, human–model interaction, and researcher formation, this article argues that disclosure documents model use, whereas responsible use requires claim-level authorial warrant: a contributor's capacity to understand, justify, and defend claims within their contribution. The article distinguishes cognitive delegation, in which model outputs shape substantive choices, interpretations, or justifications, from routine delegation, which does not ordinarily bear on the substance of judgment. Its conceptual contribution is to link the epistemic role of outputs that shape judgment to the function-appropriate evaluation required to justify decisions and retain warrant for resulting claims. It then repurposes disclosure categories as a claim-sensitive ex ante heuristic, mapping six recurrent research functions to risks, forms of oversight, and formative questions for recognising cognitive delegation and selecting function-appropriate evaluation. The article contributes to the ethics of information technology by clarifying when AI-supported scholarship remains under responsible human judgment and when cognitive delegation threatens the epistemic basis of authorship.

## Introduction

Editorial policies increasingly emphasise human responsibility and some form of disclosure, while the scope, thresholds, and purposes of these requirements remain inconsistent (Hosseini et al., 2023; Nahar et al., 2025; Resnik & Hosseini, 2026). Disagreement concerns not only when disclosure is required, but also which uses are substantively relevant and what disclosure is expected to achieve.

Where a model materially generates or modifies analytic code, produces synthetic data, or alters a research procedure, its use is methodologically relevant and should be documented. The concern is inferential: documentation and authorial warrant—the contributor's claim-level capacity to understand, justify, and defend specific claims, and thereby to take responsibility for them—answer different questions and should not be treated as interchangeable.

Disclosure can document an author's account of where and how a model was used. Such information may be methodologically relevant, but it does not establish whether the contributor retained sufficient understanding of the resulting claims and adequate oversight of the model-supported process that shaped them. A statement such as "I used model X to suggest citations" identifies an activity; a fuller account may describe claimed safeguards. Neither account, by itself, establishes how the output shaped judgment or whether authorial warrant was retained.

Existing frameworks need not explicitly equate disclosure with warrant for this limitation to matter: record completeness can support scrutiny, but cannot by itself determine whether the contributor retained sufficient understanding, applied function-appropriate evaluation, and remained able

✉ Marcin Mirończuk
marcin.mironczuk@opi.org.pl

1 AI Lab, National Information Processing Institute, Warsaw, Poland

to take responsibility for the claims at issue. Journal policies commonly require human contributors to take responsibility for AI-assisted work. On the account developed here, fulfilling that requirement depends on retaining authorial warrant for claims within one's contribution. The aim is to clarify the claim-level epistemic capacities needed to meet this existing responsibility, rather than to introduce an additional reporting requirement.

The concern is not that disclosure policies explicitly equate reporting with authorial warrant, but that reporting alone does not cultivate the epistemic and formative capacities on which responsible use depends. Where its purpose is left unclear, a disclosure requirement may orient attention toward compliance rather than toward those practices. Institutional guidance should therefore extend beyond currently salient failure modes, such as hallucinations or false citations, toward more model-agnostic principles of delegation, function-appropriate evaluation, and responsibility.

Detector-mediated approaches may also encourage model use to be understood through output-level indicators, such as whether the final text appears AI-generated or receives a high AI-likelihood score (Conroy, 2023; Hu, 2026). Even where policies do not formally treat such indicators as evidence of the extent of model use, this product-level proxy can shift attention away from the epistemic relationship between the contributor and the claims for which they provide warrant.

Collectively, recent frameworks map activities supported by generative models, distinguish disclosure thresholds, and specify degrees of human oversight (Weaver, 2024; Resnik & Hosseini, 2026; Suchikova et al., 2026; Ahmetoglu et al., 2026). Recent editorial guidance provides stage-based ex ante safeguards and advocates proportional assessment according to the function, intensity, and epistemic risk of AI assistance (Pinochet et al., 2026). What remains underdeveloped is a claim-sensitive criterion that tracks when the role of a model output in a specific claim or decision marks the transition from routine delegation to cognitive delegation and links that role to the evaluation needed to justify the relevant decisions and retain warrant for the resulting claims. The article therefore argues that disclosure documents model use, whereas responsible use requires claim-level authorial warrant, and that categories developed for disclosure can be repurposed ex ante to help researchers recognise cognitive delegation and select function-appropriate evaluation.

This theoretical perspective is informed by a selective, problem-driven synthesis of literature from research ethics, disclosure policy, human–model interaction, and researcher formation; it does not aim to provide a systematic or exhaustive review. The literature is used to reconstruct the debate, compare the functions of existing frameworks, and develop a conceptual argument about authorial warrant.

Accordingly, institutional guidance should orient researchers toward asking which functions they delegate, how outputs are evaluated, which claims or decisions those outputs shape, whether they can justify the relevant decisions, and whether they can reconstruct and defend the resulting claims without treating model outputs as sufficient justification. On this account, these capacities are cultivated through research practice, guidance, training, and supervision and are required for responsible authorship.

To illustrate its implications across research functions, the article maps six recurrent functions across the research process and relates them to context-dependent risks, forms of oversight, and formative questions. The analysis thereby extends the discussion from documenting model use to cultivating warrant and concludes by considering how this claim-sensitive account extends to increasingly automated research workflows.

## Authorial warrant

This article uses the concept of authorial warrant here in a deliberately narrow sense: as the contributor's claim-level capacity to understand, justify, and defend the claims within their contribution. Responsible authorship requires the contributor to retain authorial warrant for those claims. Responsibility for a scientific claim is substantive rather than merely nominal only where the contributor retains sufficient epistemic access to the reasons on which the claim depends. Warrant is therefore a necessary epistemic basis for taking responsibility for them, not a complete account of responsibility itself. The capacity to reconstruct the relevant reasoning is treated not as a separate component of warrant, but as a practical test of whether the contributor's understanding, justification, and capacity for defence have been retained.

Authorial warrant should not be understood as answerability for the paper as a whole. Levy (2025), drawing on Shoemaker's distinction (Shoemaker, 2011) between attributability, answerability and accountability, argues that answerability for a paper as a whole is often absent even in radically collaborative human science, and uses this to challenge the view that AI can be excluded from authorship simply because it cannot be answerable for a paper. Among the options he canvasses is requiring only local responsibility for an intellectual contribution. This article takes up and develops that option: authorial warrant is required locally for the claims shaped or supported by each delegated function, rather than for the paper as an undifferentiated whole. A human contributor may hold such local warrant for claims within a contribution; a generative model cannot hold authorial warrant because it cannot understand, justify, or defend those claims; lacking that epistemic basis—and not being the kind of agent that can bear responsibility—it cannot take responsibility for them either.

A related view of human responsibility appears in Floridi's (2025) account of "distant writing," which treats the human as a responsible designer who specifies requirements, curates model-generated text, and remains accountable for the final literary work. The contrast with research authorship is instructive: design control and final approval are not sufficient where model outputs shape scientific claims unless the contributor also retains the capacity to understand, justify, and defend those claims, and so to take responsibility for them.

Cognitive delegation occurs when the output of a delegated function is used as a basis for substantive selection, interpretation, decision-making, or justification and thereby shapes the content of a paper's claims or their support—for example, through the selection of a hypothesis, the choice of method, the interpretation of data, or the assessment of a source's credibility. It differs from routine delegation, such as formatting, language editing, or the mechanical search for references, which does not ordinarily bear on the substance of the author's judgment. For example, using a model to assemble a pool of candidate references can remain routine when the pool is not treated as exhaustive, the author independently assesses each source, and — where comprehensiveness matters — supplements it with an appropriate literature-search method. This illustrates the additional analytical resolution provided by a claim-sensitive criterion: within the same literature-orientation function, model use may remain routine delegation or become cognitive delegation depending on whether the output is used as a basis for substantive selection and judgment.

The boundary into cognitive delegation is crossed not by the choice of tool, but when the researcher uses a model output as a basis for deciding which claims, sources, methods, interpretations, or conceptual frames should count. Cognitive delegation can be permissible, but it requires oversight grounded in function-appropriate evaluation of the delegated result. Depending on the function, such evaluation may include source checking, testing, comparison with alternatives, or explicit justification of selection. These practices do not by themselves establish authorial warrant, but they provide an observable basis for assessing its epistemic dimensions—whether the contributor can understand, justify, and defend the claims shaped by the delegation. Such oversight does not require the ability to produce the delegated result independently from scratch. The relevant human contributor remains responsible for the delegated process: they understand what has been delegated and retain warrant for the claims shaped or supported by that contribution in the final work.

## Two perspectives in the debate

The debate can be organised on two levels. The first, institutional-regulatory level includes policies, guidelines, disclosure practices, detectors of machine-generated text, and studies of their implementation, use, perception, and possible behavioural effects. The second, formative level concerns researchers themselves: how working with such models affects understanding, authorship, responsibility, writing as a cognitive process, critical thinking, and the development of research competencies.

### Institutional-regulatory perspective

#### The conceptual-normative strand

The public dissemination of generative tools based on large language models, especially following the emergence of ChatGPT, intensified debates over whether and how their use should be disclosed, whether detectors should be used, and whether such tools can qualify as authors. Major editorial policies commonly exclude generative models from authorship and retain responsibility with human contributors, a position consistent with the ICMJE authorship criteria, the COPE position, and Nature's editorial policy (Hosseini et al., 2023). The concept of authorial warrant — in the sense set out above — explains why this rule is not merely administrative: a model cannot satisfy a necessary epistemic condition of responsible authorship because it cannot hold authorial warrant for claims.

The following discussion focuses on two recurring institutional-regulatory questions: how model use should be disclosed and whether model-generated text should be detected. Conroy (2023), in a Nature feature, surveys the emerging landscape—bans at some publishers, transparency requirements at others, and uneven guidelines on reporting LLM use. The feature highlights two equity-related concerns: generative models may assist researchers who do not write in English as a first language, particularly when their use is appropriately disclosed, whereas AI-detection tools may wrongly flag such prose as machine-generated. It also notes that verifying whether authors genuinely conducted the research cannot be delegated entirely to journals. Detection has drawn criticism on related but distinct grounds: Hu (2026) warns of a "style tax," whereby the prospect of detection may lead researchers to simplify their grammar and flatten their prose to avoid appearing machine-generated — and argues, in a direction this article develops, for verifiable process evidence such as documented workflows over probability scores assigned to the final text.

Alongside criticism of detection, normative and practical work developed principles, thresholds, and structured categories for disclosure. Blau et al. (2024) formulated five principles of human accountability and responsibility for the use of generative models in science: disclosure and attribution, verification, documentation of AI-generated data, ethics and equity, and continuous monitoring, oversight, and public engagement. Weaver (2024) AID Framework provides structured categories that can be included in a declaration of model use instead of the uninformative statement “we used ChatGPT.”

Resnik and Hosseini (2026) further propose a threefold distinction: disclosure should be mandatory only when model use is “intentional and substantial” — that is, when it produces evidence, analysis, or content that affects the study’s conclusions or directly shapes the text; optional for context-dependent uses such as editing, finding references, brainstorming, or feedback on existing ideas; and unnecessary for incidental uses, such as phrase suggestions or a model embedded in larger systems.

These principles and thresholds specify when and what to disclose. Other frameworks operationalise how disclosure is structured in practice by identifying delegated tasks and organising them into reporting instruments. GAIDeT (Suchikova et al., 2026) classifies delegated tasks at macro and micro levels, indicating the required degree of human oversight, while DAISY (Ahmetoglu et al., 2026), building on prior AI disclosure frameworks including AID, operationalises disclosure through a structured form-based tool that guides authors across categories of AI use. GAIDeT and DAISY substantially operationalise the reporting axis by organising what and how to disclose. Taken together with the thresholds proposed by Resnik and Hosseini, these frameworks structure what, when, and how model use should be reported.

### The empirical strand

Empirical work examines disclosure practices, policy implementation, and evaluations of model use, and three empirical observations bear on the argument developed here. First, model-related journal policies are widespread, but explicit disclosure remains rarely visible in published full texts. He and Bu (2026) found that although roughly 70% of the analysed journals had model-related policies, explicit disclosure appeared in only about 0.1% of the full-text publications examined since 2023, even as statistical indicators suggested substantially broader use. The disclosure rate nevertheless showed a slow upward trend, rising from approximately 0.01% in early 2023 to 0.43% in the first quarter of 2025. The authors suggest that policies may be moving a subset of researchers from use towards declaration, although their analysis does not establish that policies caused this increase. The findings therefore indicate a substantial gap between the prevalence of model-related policies and observable disclosure in published articles, while also showing that disclosure is increasing from a very low baseline.

Second, researchers’ judgements about acceptable model use and the conditions under which it should be disclosed vary substantially by task. A Nature survey of 5,229 academics (Kwon, 2025) found that more than 90% considered editing and translation acceptable under at least some conditions, whereas text generation and peer-review uses were more contested. More than 60% rejected having AI generate an initial review report, whereas 57% accepted its use to assist a human reviewer, for example by answering questions about a manuscript. Earlier survey work provided broader background on researchers’ uses of and attitudes towards AI (Van Noorden & Perkel, 2023), while Nahar et al. (2025) documented inconsistency in policy adoption, disclosure requirements, and restrictiveness across computer-science conferences.

Third, an experimental study bears on how task-specific disclosure may shape perceptions when evidence about verification and integration is absent. In an experiment with 402 US-based lay participants, Niszczota and Conway (2023) found that delegating parts of the research process to a language model rather than to a PhD student was judged less morally acceptable, reduced trust in the researcher to oversee future projects, and lowered expectations of the accuracy and quality of the resulting work. The delegated function was specified, but participants received no information about how the output was verified or integrated. The study therefore concerns non-expert perceptions rather than the judgements of researchers, editors or reviewers, whom the authors acknowledge might respond differently. It does not show that model use lowers research quality. Rather, it suggests that, in the absence of evidence about verification and integration, disclosure focused on the fact of model use may trigger negative judgements about the researcher and the expected quality of the work before readers can assess how the output was verified and integrated. This motivates attention to the relationship between model use, human evaluation, and the resulting claims.

### Conclusions from the synthesis

Two strands emerge from this literature. On the conceptual-normative side, major editorial policies commonly exclude generative models from authorship and retain responsibility with human contributors, while reporting

frameworks—AID, GAIDeT, DAISY, and Resnik and Hosseini's tripartite distinction—organise the scope of use, delegated tasks, reporting thresholds, and human oversight. Detection of the final text, by contrast, remains a weak basis for control because it is unreliable and liable to impose a "style tax." On the empirical side, researchers' judgements about acceptable model use and its conditions vary substantially by task: editing and translation are widely accepted under at least some conditions, whereas content generation and model use in peer review remain more contested. A substantial gap separates the prevalence of model-related policies from the rarity of explicit declarations in published full texts. Taken together, these strands provide detailed accounts of reporting requirements, categories, and observed disclosure practices. What these strands do not provide is a claim-sensitive formative criterion for determining whether model use leaves the researcher able to understand, justify, and defend the resulting claims. It is to that dimension that the discussion now turns.

## Formative perspective

### The conceptual-normative strand

In parallel with the institutional debate on disclosure, a formative strand can be reconstructed from work on scientific understanding, cognitive agency, critical interaction with models, and the development of research competencies. Early discussions surveyed both the opportunities and risks of these tools for science (Stokel-Walker & Van Noorden, 2023), while Messeri & Crockett (2024) identify a deeper epistemic problem: AI-assisted scientific practices may create "illusions of understanding," in which researchers believe that they understand more than they actually do, while the emergence of narrower methods, questions, and viewpoints becomes harder to recognise. For the present argument, the central risk is therefore not model use itself, but the substitution of plausible output for the researcher's own understanding and justification.

A constructive response requires an adequate conception of the tool and a mode of interaction that preserves judgement. Shanahan (2024) cautions against uncritically attributing human-like epistemic capacities to language models. Sarkar (2024) develops the corresponding positive proposal: the model should function as a provocateur rather than as a secretary, exposing assumptions, generating counterarguments, identifying weaknesses, and supporting reflection instead of producing finished reasoning on the researcher's behalf. Broader debates over collaboration foreground disagreement about the roles models may perform and about who should retain responsibility for the direction and evaluation of scientific work (Binz et al., 2025). Read alongside work on writing and expertise, this debate acquires formative significance: literature synthesis, study design, argumentation, and writing are not only productive tasks but practices through which scientific competence is formed and maintained (Editorial, 2025; Lin and Sohail, 2026).

Lin and Sohail (2026) systematise this formative response through three "AI meta-skills": strategic direction, critical discernment, and systematic calibration. These capacities concern deciding whether and when to use a model, critically evaluating its outputs, and validating whether they should be accepted, revised, or rejected. Their framework shifts attention from general warnings about dependence to the competencies needed to use models without weakening intellectual autonomy.

Read together, these works motivate a move from asking merely whether a model was used to examining how its outputs entered the researcher's judgement. In the terms developed here, when an output shapes evidence selection, methods, interpretations, or justifications, the relevant issue is whether the researcher retains cognitive agency, can reconstruct the resulting reasoning, and can justify and defend the claims it supports. This motivates the distinction developed here between routine and cognitive delegation and the need to link delegated functions to function-appropriate evaluation.

### The empirical strand

The empirical literature provides indirect evidence about the formative consequences of model use through studies of writing, learning, retention, ownership, and subsequent independent performance. These studies do not examine the research practice of experienced scientists and should not be treated as direct evidence about research quality. They are better read as task- and population-limited signals concerning activities through which relevant competencies are developed or exercised — activities that generative models can partially perform on a user's behalf. For the present argument, their relevance lies primarily in researcher formation and, more cautiously, in the calibration of model use needed to preserve authorial warrant.

Two studies provide limited signals about memory, ownership, and the possible costs of reduced effortful engagement. In a small preprint study of essay writing, Kosmyna et al. (2025) found that LLM users reported weaker and more variable ownership of their texts and showed lower immediate ability to quote what they had written than participants in the comparison conditions. A fourth-session comparison involving only 18 participants suggested better recall among those who wrote without a model before using one than among those who began with model assistance. These findings are treated here as task-specific and exploratory;

the reported EEG differences are not used as evidence about scientific competence or research quality. Barcaui (2025), in a randomised study of undergraduate learning, found that students given unrestricted access to ChatGPT scored lower on a delayed retention test after 45 days than those using traditional study methods. The result is consistent with accounts of cognitive offloading and desirable difficulties, although the study did not directly measure encoding processes or learning strategies. Together, the studies provide a limited formative signal that, in these tasks, some forms of model assistance were associated with weaker retention, ownership, or immediate recall of the produced text.

Wong and Qiu (2026) provide a complementary result concerning the mode and timing of assistance. In a creative problem-solving task, free ChatGPT use improved creative performance while assistance was available, but this advantage did not persist when participants subsequently worked independently. Participants following a "think first, ChatGPT later" procedure — generating their own ideas before using the model to improve, develop, and evaluate them — performed best on the later independent task. Within this setting, the result supports a distinction between model use that substitutes for an initial independent attempt and use that follows and develops one. It does not establish that the same sequencing effect holds across other research or learning tasks.

Taken together, these studies do not justify a general claim that generative models impair cognition. They support a narrower proposition: formative outcomes may be sensitive to when and how model assistance enters the task and whether the user first makes an independent attempt. Within the present account, this motivates treating timing and mode, alongside the epistemic role of the output, as relevant to cognitive delegation and evaluation.

### Conclusions from the synthesis

Taken together, the conceptual arguments and empirical signals illustrate the importance of a formative dimension in model-supported work. Model use should preserve the researcher's capacity to understand, justify, and defend the claims shaped through that use, and thereby to take responsibility for them, with reconstruction of the relevant reasoning serving as a practical test of whether the epistemic dimensions of warrant have been retained. The empirical studies do not show that generative models generally impair cognition. They suggest more narrowly that formative outcomes may depend on when and how assistance enters a task, whether it substitutes for an initial independent attempt, and whether assisted performance persists once the model is removed. The literature therefore directs attention beyond the fact of model use towards cognitive agency, critical evaluation, and the mode and timing of interaction. It offers constructive principles, including the use of models for critique rather than substitution and the meta-skills identified above. What it does not by itself provide is a claim-sensitive instrument that connects a particular delegated research function to the form of evaluation required for the researcher to retain warrant for the resulting claim. The formative dimension is thus clearly articulated, but remains insufficiently connected to specific acts of delegation and claim-level responsibility.

## Synthesis and gap

Across the texts discussed, disclosure, verification, and human responsibility remain central, and several frameworks now organise types of use, disclosure thresholds, and degrees of human oversight. Recent editorial guidance goes further by providing stage-based ex ante safeguards and advocating proportional assessment according to the function, intensity, and epistemic risk of AI assistance (Pinochet et al., 2026). What remains underdeveloped is not ex ante guidance as such, but the claim-sensitive repurposing of these categories: a criterion that tracks how a model output enters a particular claim or decision, when that role marks the transition from routine delegation to cognitive delegation, and what function-appropriate evaluation is needed to justify the relevant decisions and retain warrant for the resulting claims.

Why is this a problem for research integrity, and not merely for terminology? When cognitive delegation goes unrecognised, a researcher may assume responsibility for claims whose underlying reasoning they can no longer reconstruct and which they can no longer justify or defend without treating model outputs as sufficient justification. The problem therefore concerns the epistemic basis of responsible authorship and the formation of the researcher—their competence, understanding, judgment, and capacity to take responsibility for the claims made under their name. Stage- or task-based categories cannot by themselves identify this loss of control, because within the same research function model use may remain routine delegation or become cognitive delegation depending on whether the output is used as a basis for substantive selection, interpretation, or judgment. What matters is therefore not merely the fact or stage of model use, but the function delegated and the epistemic role of the output. Timing and mode may also matter, as suggested by studies based on the principle "think first, ChatGPT later."

The gap that emerges from the analysis is therefore not the absence of disclosure mechanisms or ex ante safeguards, but the absence of a sufficiently explicit bridge between disclosure categories and the claim-sensitive formative practices needed to preserve authorial warrant. The existing frameworks—AID, GAIDeT, DAISY, and Resnik and Hosseini's tripartite distinction—retain their documentary and methodological value; for the purpose developed here, their categories can also be repurposed as material for guidance, training, and reflective practice, in addition to their reporting and methodological uses. A disclosure record, however complete, neither tells a researcher how to work nor cultivates the competence on which responsible use depends. The heuristic criterion developed here is intended to distinguish routine delegation from cognitive delegation and, within cognitive delegation, to distinguish uses after which the author remains able to justify the relevant decisions, reconstruct the relevant reasoning, and defend the resulting claims from uses after which they cannot do so without treating model outputs as sufficient justification. The issue is not to prohibit the use of models, nor to perfect the disclosure form, but to enable a mode of use that preserves this capacity. What this asks of the author is that they retain authorial warrant; what it asks of institutions is that they cultivate it.

## A claim-sensitive map of delegated research functions

The following map is not intended as a complete taxonomy of generative-model use. It is a heuristic synthesis drawn from existing disclosure frameworks, especially AID and GAIDeT, and from recurrent functions across research practice: writing, literature orientation, analysis, interpretation, methodological choice, and conceptual framing. Its purpose is narrower: to show how attention can move from naming a tool to identifying the function delegated to it and the epistemic role its output came to play, the associated risks, an indicative form of oversight, and a formative question the researcher should be able to answer.

The map does not assign a fixed epistemic status to any function: model use within the same function may remain routine delegation or become cognitive delegation depending on how its output enters a specific claim or decision.

For that purpose, six recurrent functions are useful as a starting point, not as an exhaustive classification. Each function is associated with a different type of risk, calls for a different form of oversight, and poses a different formative question for the researcher. The point is not to report model use as an abstract fact, but to clarify, for the researcher, where the model entered the research process, what role it played in the process, how its output was evaluated, and whether the relevant contributor can justify the relevant decisions and retain authorial warrant for the claims shaped by the delegated function.

A reflective practice built around delegation, function-appropriate evaluation, and authorial warrant would not by itself guarantee responsible use. It could, however, help orient researchers toward the questions that matter for responsible cognitive work. In this sense the guidance and tools that structure such a practice are part of the behavioural infrastructure of research no less than disclosure forms, but they are better suited to shaping researchers' understanding of authorship, delegation, and responsibility in a reflective rather than merely administrative direction.

- Language and editing require checking whether a change in style has changed the meaning.
- Search and orientation in the field require verification of every citation against the cited source, and consultation of primary evidence where the claim requires it; bibliographic hallucinations are a typical failure mode of language models.
- Programming and analysis require code review, tests and checks of results against reference data.
- Interpretation requires that the author be able to justify a reading of the data without treating the model's proposal as sufficient justification.
- Methodological choice requires that the procedure can be defended before a reviewer; methodology shapes results in ways that are often invisible.
- Conceptual framing poses distinctive conditions for retaining authorial warrant because it can shape the research question itself, determine what counts as relevant evidence, and constrain the range of later interpretations.

| Delegated function | Typical risk | Indicative oversight | Formative question |
|---|---|---|---|
| Language and editing | shift in meaning | comparison of versions | Do I understand and accept the change in meaning? |
| Search and orientation in the field | hallucinations, omissions | checking cited sources; consulting primary evidence where required | Do I myself know why this literature matters? |
| Programming and analysis | faulty code, faulty assumptions | tests, review, reference data | Can I explain the procedure? |

| Delegated function | Typical risk | Indicative oversight | Formative question |
|---|---|---|---|
| Interpretation | adopting someone else's frame | consideration of alternative interpretations | Can I reconstruct, justify, and defend this interpretation without treating model outputs as sufficient justification? |
| Methodological choice | hidden shift in results | a defensible account of the method choice | Can I justify why this method is appropriate for the research question? |
| Conceptual framing | taking over the research question | reconstructing the assumptions oneself | Is this still my research question? |

## What a researcher should be able to answer throughout the research process

The following questions are not proposed as a new mandatory form to be completed at submission. They provide a starting framework for the researcher's own use — questions one should be able to answer about one's own process, ideally set as a frame before working with a model rather than answered only retrospectively. Used this way, they shift the work from a retrospective account to a habit of delegating only in ways that leave the researcher able to justify the relevant decisions, reconstruct the relevant reasoning, and defend the resulting claims:

- What functions did I delegate to the model in this project? Language, search, analysis, interpretation, methodology, conceptual framing?
- Which claims or decisions did model outputs shape, and how?
- Which model outputs did I accept, which did I modify, which did I reject – and why?
- How did I evaluate each output in a way appropriate to the function it performed?
- For the central claims, can I reconstruct the reasoning, identify the evidence, explain the analytical or interpretive steps, and state where my own judgment overrode or confirmed the model's output?

The sequence culminates in the final question, which should not be treated as an invitation to a generic assurance. It asks the researcher to provide a reconstructable account offering evidence that the epistemic dimensions of authorial warrant have been retained.

## Repurposing disclosure categories to cultivate warrant

The questions above describe what a researcher should be able to answer about their own work. They are addressed primarily to the researcher, although supervisors and examiners may use them to structure discussion of a contribution. The remaining question is how journals, institutions, and tools can be designed so that researchers are drawn toward asking them—not by attempting to infer warrant from a disclosure, but by shaping the environment in which research is done.

The point is not that reviewers should read a disclosure and judge from it whether an author is responsible. Authorial warrant is not established by a submission field; its epistemic dimensions are more meaningfully tested when the author must reconstruct the reasoning behind the claims and justify and defend the claims themselves—in the defence of a thesis, in the response to review, or when a colleague asks why this method and not another. An author who delegated without retaining understanding lacks the epistemic basis for such a defence, even if that deficiency will not always be detectable in a single exchange.

For journals and institutions, then, the most useful contribution to preserving authorial warrant is not simply a richer disclosure field, but guidance that repurposes disclosure categories for planning, training, and reflective practice. Their role is to articulate what responsible use looks like in a given discipline, to teach it, and to point researchers toward established good practice—much as reporting standards such as PRISMA or STROBE can make researchers aware ex ante of which elements of the process should be planned and documented. Where an institution wants a declaration for the purpose of addressing authorial responsibility—rather than methodological provenance—a more epistemically meaningful form is a confirmation that the author followed the practices the community regards as sound: an attestation of process, not merely an inventory of delegation. The attestation is offered as a way of reorienting declarations already requested, not as a case for introducing an additional requirement.

Such an attestation is no more proof against bad faith than a conventional disclosure: an author who would misreport tool use could also affirm practices they did not follow. Its value lies in directing attention toward the practice itself, not in establishing authorial warrant at submission. The relevant distinction is therefore not merely whether an institution requires a declaration, but whether the declaration points toward the researcher's craft or merely inventories model use.

For doctoral students and their supervisors, conversations about generative models should not be reduced to whether the model was used or whether it touched the text. They should lead to questions about which claims the student can justify and defend, and whether they can reconstruct the reasoning behind them without treating model outputs as sufficient justification, and how supervision can develop that capacity. Competence in working with these models should be taught in the same way as statistical methods or research ethics: not as a set of prohibitions, but as capacities to recognise when delegation becomes cognitive, apply function-appropriate evaluation, and identify what should not be delegated when the researcher's own development is at stake. The primary aim is formative, although assessment can both support learning and help judge whether the contributor meets the relevant expectations of understanding and justification.

More broadly, cultivating these capacities calls for repeated practice in evaluating outputs, comparing alternatives, and explaining why a model's proposal was accepted, modified, or rejected. For those still acquiring disciplinary competence, an independent attempt before model assistance may be especially valuable when the aim is to develop the reasoning that will later be delegated; for experienced researchers, the task is to recalibrate delegation and evaluation within changing workflows. Live questioning without model assistance offers one possible way to probe the understanding and justification underlying a contribution. It should address evidence, assumptions, methodological choices, and limitations, without requiring independent reproduction of every delegated result or treating a single exchange as conclusive. In educational settings, the ability to explain and justify the reasoning underlying one's work can serve as an explicit assessment criterion at successive stages of a project. In publication, scrutiny of specific claims can prompt further evaluation, revision, or qualification where justification is inadequate. These formative and assessment applications are proposals derived from the conceptual account and have not been empirically validated as ways of cultivating or assessing authorial warrant; their effectiveness, feasibility, and fairness require evaluation across disciplines and career stages.

For tool designers: build productive friction. This does not mean UX obstacles, but moments built into the workflow that ask the researcher to identify a source, formulate uncertainty, or list rejected alternatives. A good system supporting research work should leave the researcher with a better understanding of the problem, not merely with finished text.

## Future perspective: automated research and the preservation of authorial warrant

The problem of preserving authorial warrant does not end with writing supported by generative models. Emerging research-automation systems extend it from discrete model uses within particular research functions to increasingly integrated workflows encompassing ideation, experimental design, data analysis, manuscript writing, and preliminary evaluation. Lu et al. (2026) describe "The AI Scientist" as a process leading from idea to manuscript and automated evaluation, encompassing activities that have traditionally formed part of the core of scientific work, at least in areas susceptible to computational automation. Musslick et al. (2025) frame this development more broadly as the automation of scientific practice and a transformation of the researcher's role within increasingly automated scientific systems.

As multiple functions are combined within end-to-end workflows, identifying only the stages at which AI was used becomes less informative. The relevant questions are where human judgment and calibration remained, how model-generated outputs shaped particular claims or decisions, whether function-appropriate evaluation was applied, and whether the researcher retained authorial warrant for the resulting claims. End-to-end automation therefore strengthens rather than removes the need for a claim-sensitive account: responsibility cannot be located merely by naming the automated system or research stage, but must be traced through the functions delegated, the decisions shaped, and the claims for which the human contributor retains authorial warrant.

## Conclusions

The questions typically asked in model-use disclosure are insufficient for assessing responsible use. Knowing whether, where, and how a model was used can provide important documentary and methodological information, but it does not establish how model outputs shaped particular claims or decisions, whether function-appropriate evaluation was applied, or whether the contributor retained authorial warrant for the resulting claims. The relevant questions therefore concern which functions were delegated, when model outputs became a basis for substantive selection or judgment, which decisions the researcher can justify, and whether they can reconstruct the relevant reasoning and justify and defend the resulting claims without treating model outputs as sufficient justification.

More precise disclosure taxonomies alone cannot answer these questions. Existing frameworks—including

GAIDeT, AID, DAISY, and Resnik and Hosseini's tripartite distinction—help specify what to report, when reporting is required, and how different forms of model use can be described. Their categories can also be repurposed ex ante as scaffolds for guidance, training, and reflective practice. In a world in which increasingly integrated systems can perform multiple research functions on the researcher's behalf, the shift developed in this article—from documenting model use to tracing delegated functions, cognitive delegation, and retained warrant—becomes more important rather than less.

The issue is therefore not only where a model was used, but where human judgment and calibration remained and whether the researcher retained the capacities required for responsible authorship. The ability to recognise when delegated functions become cognitive delegation, apply function-appropriate evaluation, justify the relevant decisions, and retain authorial warrant for the resulting claims belongs to the researcher's craft. Journals and institutions should cultivate these capacities through disciplinary guidance, training, supervision, and tools that repurpose disclosure categories for formative use. Where model use is methodologically material, it should still be reported as part of the method. For the purpose of authorial responsibility, however, disclosure can document an author's account of relevant practices but cannot establish their adequacy or substitute for claim-level authorial warrant.

**Author contributions** The author has contributed extensively to the work presented in this paper and satisfies the criteria for authorship. The contributions, described by CRediT roles, include Conceptualization, Investigation, Validation, Methodology, Project administration, Writing—original draft.

**Funding** The author(s) received no financial support for the research, authorship, and/or publication of this article.

**Data availability** No datasets were generated or analysed during the current study.

## Declarations

**Conflicting interests** The author(s) declared no potential conflicts of interest with respect to the research, authorship, and/or publication of this article.

**AI Statement** In accordance with journal policy, generative models were used for language editing, translation support, structural feedback, and critical revision support, including eliciting counterarguments, identifying implicit assumptions, and testing the coherence of the argument. All model-supported claims, citations, interpretations, and formulations were evaluated by the author; factual claims and citations were verified, and all final formulations were reviewed and revised by the author, who remains fully responsible for the submitted manuscript.

## References

Ahmetoglu, Y., Constantinides, M., & Cox, A. (2026). *AI Disclosure with DAISY*.

Barcaui, A. (2025). ChatGPT as a cognitive crutch: Evidence from a randomized controlled trial on knowledge retention. *Social Sciences & Humanities Open*, *12*, 102287. https://doi.org/10.1016/j.ssaho.2025.102287

Binz, M., Alaniz, S., Roskies, A., Aczel, B. A., Bergstrom, C. T., Allen, C., Schad, D., Wulff, D., West, J. D., Zhang, Q., Shiffrin, R. M., Gershman, S. J., Popov, V., Bender, E. M., Marelli, M., Botvinick, M. M., Akata, Z., & Schulz, E. (2025). How should the advancement of large language models affect the practice of science? *Proceedings of the National Academy of Sciences*, *122*(5). https://doi.org/10.1073/pnas.2401227121

Blau, W., Cerf, V. G., Enriquez, J., Francisco, J. S., Gasser, U., Gray, M. L., Greaves, M., Grosz, B. J., Jamieson, K. H., Haug, G. H., Hennessy, J. L., Horvitz, E., Kaiser, D. I., London, A. J., Lovell-Badge, R., McNutt, M. K., Minow, M., Mitchell, T. M., & Ness, S.… Witherell, M. (2024). Protecting scientific integrity in an age of generative AI. *Proceedings of the National Academy of Sciences*, *121*(22). https://doi.org/10.1073/pnas.2407886121

Conroy, G. (2023). How ChatGPT and other AI tools could disrupt scientific publishing. *Nature*, *622*(7982), 234–236. https://doi.org/10.1038/d41586-023-03144-w

Floridi, L. (2025). Distant Writing: Literary Production in the Age of Artificial Intelligence. *Minds and Machines*, *35*(3), 30. https://doi.org/10.1007/s11023-025-09732-1

He, Y., & Bu, Y. (2026). Academic journals' AI policies fail to curb the surge in AI-assisted academic writing. *Proceedings of the National Academy of Sciences*, *123*(9). https://doi.org/10.1073/pnas.2526734123

Hosseini, M., Resnik, D. B., & Holmes, K. (2023). The ethics of disclosing the use of artificial intelligence tools in writing scholarly manuscripts. *Research Ethics*, *19*(4), 449–465. https://doi.org/10.1177/17470161231180449

Hu, B. (2026). Why artificial intelligence detectors could penalize academic writing. *Nature Human Behaviour*, *10*(3), 439–439. https://doi.org/10.1038/s41562-026-02420-9

Kosmyna, N., Hauptmann, E., Yuan, Y. T., Situ, J., Liao, X. H., Beresnitzky, A. V., Braunstein, I., & Maes, P. (2025). *Your Brain on ChatGPT: Accumulation of Cognitive Debt when Using an AI Assistant for Essay Writing Task*.

Kwon, D. (2025). Is it OK for AI to write science papers? Nature survey shows researchers are split. *Nature*, *641*(8063), 574–578. https://doi.org/10.1038/d41586-025-01463-8

Levy, N. (2025). Responsibility is not required for authorship. *Journal of Medical Ethics*, *51*(4), 230–232. https://doi.org/10.1136/jme-2024-109912

Lin, Z., & Sohail, A. (2026). Recalibrating academic expertise in the age of generative AI. *Patterns*, *7*(1). https://doi.org/10.1016/j.patter.2025.101473

Lu, C., Lu, C., Lange, R. T., Yamada, Y., Hu, S., Foerster, J., Ha, D., & Clune, J. (2026). Towards end-to-end automation of AI research. *Nature*, *651*(8107), 914–919. https://doi.org/10.1038/s41586-026-10265-5

Messeri, L., & Crockett, M. J. (2024). Artificial intelligence and illusions of understanding in scientific research. *Nature*, *627*(8002), 49–58. https://doi.org/10.1038/s41586-024-07146-0

Musslick, S., Bartlett, L. K., Chandramouli, S. H., Dubova, M., Gobet, F., Griffiths, T. L., Hullman, J., King, R. D., Kutz, J. N., Lucas, C. G., Mahesh, S., Pestilli, F., Sloman, S. J., & Holmes, W. R. (2025). Automating the practice of science: Opportunities, challenges, and implications. *Proceedings of the National Academy of Sciences*, *122*(5). https://doi.org/10.1073/pnas.2401238121

Nahar, M., Lee, S., Guillen, R., & Lee, D. (2025). Generative Artificial Intelligence Policies under the Microscope. *Communications of the ACM*, *68*(7), 29–33. https://doi.org/10.1145/3720536

Niszczota, P., & Conway, P. (2023). Judgements of research co-created by generative AI: experimental evidence. *Economics and Business Review*, *9*(2), 17–43. https://doi.org/10.18559/ebr.2023.2.744

Pinochet, L. H. C., Hamza, M., K., & Dwivedi, Y. K. (2026). Editorial: Generative AI, human authorship and the transformation of scholarly communication. *Revista de Gestão*, *33*, 31–40. https://doi.org/10.1108/REGE-04-2026-0058

Resnik, D. B., & Hosseini, M. (2026). Disclosing artificial intelligence use in scientific research and publication: When should disclosure be mandatory, optional, or unnecessary? *Accountability in Research*, *33*(2). https://doi.org/10.1080/08989621.2025.2481949

Sarkar, A. (2024). AI Should Challenge, Not Obey. *Communications of the ACM*, *67*(10), 18–21. https://doi.org/10.1145/3649404

Shanahan, M. (2024). Talking about Large Language Models. *Communications of the ACM*, *67*(2), 68–79. https://doi.org/10.1145/3624724

Shoemaker, D. (2011). Attributability, Answerability, and Accountability: Toward a Wider Theory of Moral Responsibility. *Ethics*, *121*(3), 602–632. https://doi.org/10.1086/659003

Stokel-Walker, C., & Van Noorden, R. (2023). What ChatGPT and generative AI mean for science. *Nature*, *614*(7947), 214–216. https://doi.org/10.1038/d41586-023-00340-6

Suchikova, Y., Tsybuliak, N., Teixeira da Silva, J. A., & Nazarovets, S. (2026). GAIDeT (Generative AI Delegation Taxonomy): A taxonomy for humans to delegate tasks to generative artificial intelligence in scientific research and publishing. *Accountability in Research*, *33*(3). https://doi.org/10.1080/08989621.2025.2544331

Van Noorden, R., & Perkel, J. M. (2023). AI and science: what 1,600 researchers think. *Nature*, *621*(7980), 672–675. https://doi.org/10.1038/d41586-023-02980-0

Weaver, K. (2024). The Artificial Intelligence Disclosure (AID) Framework: An Introduction. *College & Research Libraries News*, *85*(10). https://doi.org/10.5860/crln.85.10.407

Wong, S. S. H., & Qiu, S. X. (2026). Think First, ChatGPT Later: Guiding Human–AI Collaboration for Learning Gains in Independent Human Creativity. *Educational Psychology Review*, *38*(1), 45. https://doi.org/10.1007/s10648-026-10118-7

Writing is thinking (2025). *Nature Reviews Bioengineering*, 3(6), 431–431. https://doi.org/10.1038/s44222-025-00323-4.